%% file: cdf_psi_pol_prl_r3.tex
\documentstyle[aps,epsfig,preprint]{revtex}  % epsfig,preprint for Draft

\newcommand{\jp}{J/\psi}
\newcommand{\pp}{\psi(2S)}
\newcommand{\cts}{\cos  \theta^{\ast}}

\newcommand{\ptj}{P_{T}^{\jp}}
\newcommand{\g}{{\mathrm GeV}\!/c}
\newcommand{\ts}{\theta^\ast}
\newcommand{\aP}{\alpha_P}
\newcommand{\ab}{\alpha_B}

\def\lessim{\mathrel {\vcenter {\baselineskip 0pt \kern 0pt
\hbox{$<$} \kern 0pt \hbox{$\sim$} }}}
\def\gessim{\mathrel {\vcenter {\baselineskip 0pt \kern 0pt
\hbox{$>$} \kern 0pt \hbox{$\sim$} }}}

\def\GeVc{GeV$\!/c$}

\def\pbarp{$p\bar{p}$}

\def\etal{{\em et al.}}

\begin{document}

\draft
\preprint{}
\title{Measurement of $\jp$ and $\pp$ polarization in $p\bar{p}$
 collisions at $\sqrt{s} = 1.8$ TeV}
\date{\today}
\maketitle

\input{cdfalist.tex}

\pagebreak

\begin{abstract}
We have measured the polarization of $\jp$ and $\pp$ mesons
produced in $p\bar{p}$ collisions at $\sqrt{s} = 1.8$ TeV, using
data collected at CDF during 1992-95.
 The polarization of promptly produced $\jp$ [$\pp$] mesons
is isolated from those produced in $B$-hadron decay, and measured over
the kinematic range $ 4[5.5] < P_T < 20 \ \g$ and $|y| < 0.6$.
For $P_T \gessim 12$ $\g$ we do not observe  significant polarization 
in the prompt component.
\end{abstract}

% 13.20.Gd Decays of J/psi, Upsilon, and other quarkonia
% 13.85.Qk Inclusive production with identified leptons, photons, or other nonhadronic particles
% 13.88.+e Polarization in interactions and scattering
% 14.40.Gx Mesons with S=C=B=0, mass > 2.5 GeV (including quarkonia)

%%%%%%%%%%%%%% Introduction %%%%%%%%%%%%%%%%%%%%%%%%

The production of heavy quarkonia states, $c\bar{c}$ and $b\bar{b}$, 
provides a useful system for the study of Quantum Chromodynamics (QCD),
 as it involves
both perturbative and nonperturbative energy scales. In $p\bar{p}$ 
collisions, charmonium production occurs through three mechanisms: direct 
production, the decay of heavier charmonia, and 
the decay of $b$-flavored hadrons. The first two mechanisms are collectively
known as ``prompt'' because they are observed to occur at the $p\bar{p}$
interaction point.

The Collider Detector at Fermilab (CDF) collaboration 
previously reported results on the production 
of  $\jp$ and $\pp$ mesons \cite{cdfpsi,cdfchi}.
The measured cross sections for direct production were on the order of 
50 times larger than predicted by the Color Singlet Model (CSM)
 \cite{csmfrag}. However, calculations based on the Nonrelativistic QCD
(NRQCD) factorization formalism \cite{nrqcd,chowise}
 are able to account for the
 observed cross sections by including color octet production mechanisms.
This leads to the prediction that directly produced $\psi$
 mesons 
 will be increasingly transversely polarized at high
$P_T$ \cite{chowise,benkram,braaten}. (In this Letter we
 use $\psi$ to denote either $\jp$ or $\pp$ mesons.) 
This is because the production of $\psi$ mesons with $P_T \gg M_\psi$
is dominated by gluon fragmentation. It is predicted that
 the gluon's transverse polarization
is preserved  as the $c\bar{c}$ pair evolves
into a bound state $\psi$ meson. On the other hand, the Color Evaporation
Model predicts an absence of polarization \cite{cem}.
In this Letter, we report on measurements of the polarization of
promptly produced $\psi$  mesons at CDF. Our analysis also
yields as a byproduct the effective polarization of
the $\psi$ mesons produced in $B$-hadron decays.  

%%%%%%%%%%%%%% Detector and Dataset %%%%%%%%%%%%%%%%%%%%

CDF is a multi-purpose detector
designed to study high energy \pbarp\ collisions produced by
the Tevatron~\cite{ref: CDF Detector}.
The CDF coordinate system is defined
with the $z$\ axis along the proton beam direction.
  The polar angle $\theta$\ is defined relative to the
$z$\ axis, $r$\ is the perpendicular radius from this axis, and
$\phi$\ is the azimuthal angle.  Pseudorapidity is defined as
$\eta\equiv-\ln[\tan(\theta/2)]$.  
Three charged-particle tracking detectors
immersed in a 1.4~T solenoidal magnetic field surround the beamline.
This tracking system is
contained within a calorimeter, while drift chambers
outside the calorimeter identify muon candidates.

The innermost tracking device is a four-layer silicon microstrip detector 
(SVX) located at radii between 2.9 and 7.9 cm from the beam
axis.  The SVX is surrounded by a set of time projection chambers
(VTX) extending out to a radius of 22
cm.  An 84 layer cylindrical drift chamber (CTC) measures the particle
trajectories in the region between 30 and 132 cm  from the
beam.  This tracking system has high efficiency for detecting charged
particles with momentum transverse to the beam $P_T > 0.4$~\GeVc\
and $|\eta|\lessim1.1$.  Together, the CTC and SVX measure charged
particle transverse momenta with a precision of $\sigma_{P_T}/P_T =
0.007 \oplus 0.001 \cdot P_T$\ (with $P_T$ in \GeVc). 
The impact parameter resolution is $\sigma_{d} = (13 + 40/P_T) \ \mu$m
for tracks with SVX and CTC information.

The central muon detection system consists of four layers of planar drift
chambers separated from the interaction point by five
interaction lengths of material.  
This system detects muons with $P_T \gessim 1.4$\
\GeVc\ and $|\eta| \lessim 0.6$.  
Dimuon candidates used in this analysis are collected using a
 three-level $\mu^+\mu^-$ trigger.
The first level trigger requires that two candidates be observed in
the muon chambers.  For each muon candidate 
the first level trigger efficiency rises from
$\sim$40\% at $P_T=1.5$~\GeVc\ to $\sim$93\% for muons with
$P_T>3.0$~\GeVc.
The second level trigger requires one or more
charged particle tracks in the CTC, reconstructed using the central
fast track processor (CFT).  The CFT performs a partial 
reconstruction of all charged tracks with $P_T$ above $\sim$2~\GeVc. Muon
 candidates found by the first
level trigger are required to match a CFT track 
within 15 degrees in azimuth.
The third level trigger performs three-dimensional track reconstruction
and accepts dimuon masses in a broad window around the $\jp$ and $\pp$
masses.

The data used in this study correspond to an integrated luminosity
of 110 pb$^{-1}$ and were collected between 1992 and
1995. Following the online data collection, additional requirements are made 
offline to identify the signals and to reduce the backgrounds. 
To identify muon candidates and reduce the rate from sources
such as $\pi/K$ meson decay-in-flight, we require that each track
observed in the muon chambers be associated with a matching CTC
track. These matches are required to pass a maximum $\chi^2$ cut of 9 and
 12 (for 1 degree of freedom) in the $\phi$ and $z$ views respectively. 
Also, we require $P_T$ greater than about 2~\GeVc\ for each
muon candidate. 
This requirement ensures that the muon trigger
and reconstruction efficiencies are well understood, to
avoid biases in the decay angular distributions of the
charmonia states studied below.

%%%%%%%%%%%%%% Method %%%%%%%%%%%%%%%%%%%%%%%%%%%%%%%%%%%%%

 The measurement of the polarization of $\psi$ mesons is
 made by analyzing their decays to $\mu^{+}\mu^{-}$ in the helicity basis,
 in which the spin quantization axis lies along 
 the $\psi$ direction in the $p\bar{p}$ center-of-mass (lab) frame.
 We define $\ts$ as the
 angle between the $\mu^{+}$ direction in the $\psi$ rest frame
 and the $\psi$ direction in the lab frame.
 The normalized angular distribution $I(\cts)$ is given by

\begin{equation}
  I(\cts) = \frac{3}{2(\alpha+3)}(1+\alpha \ \cos^2\ts)
 \label{Itheta}
\end{equation}
 Unpolarized $\psi$ mesons have $\alpha=0$ whereas $\alpha=+1$ or $-1$
 correspond to fully transverse or longitudinal polarizations respectively.
 Experimentally, the acceptance is severely reduced
 as $|\cts|$ approaches 1, due to
 the $P_{T}$ cuts on the muons. Our method for determining $\alpha$
 is to fit the observed distributions of $\cts$ to
 distributions derived from simulated $\psi \rightarrow \mu^+ \mu^-$ decays.
The Monte Carlo simulation accounts for the geometric and kinematic
acceptance of the detector as well as the reconstruction efficiency
as a function of $\cts$.
                             
In order to extract the polarization parameter $\alpha$ for promptly
produced $\psi$ mesons, we separate the prompt component from
the $B$-decay component using the proper decay length of each event.
For $\psi$ candidates with one or both muons reconstructed in the SVX
(the SVX sample), we define a vector pointing from the $p\bar{p}$
collision point
to the $\psi$ decay vertex. The transverse decay length $L_{xy}$ is then 
defined as the
projection of this vector onto the $\psi$  transverse momentum.
  The proper decay length $ct$ is related
to the transverse decay length by
        $ct = (M_{\psi} L_{xy})/(F_{corr}^{\psi} P_T^{\psi})$,
where $M_{\psi}$ is the  $\psi$ mass.
Here $F_{corr}^{\psi}$ is a correction factor
  obtained from Monte Carlo studies 
 \cite{J/psi_life}, which accounts for the fact that we are using the 
 $\psi$ $P_T$ instead of the $B$-hadron $P_T$. 
Prompt events have $ct$ consistent with zero whereas $B$-decays have
 an exponential $ct$ distribution; the detector resolution smears the
 $ct$ distribution.
We fit the $ct$  distribution to obtain the relative
fractions of prompt and $B$-decay production.
 Details of this fitting procedure are given in \cite{J/psi_life}. 
 The measured fraction of $\jp$ mesons which come from $B$-hadron decay 
 increases from $(13.0\pm0.3)\%$ at $\ptj = 4 \ \g$ to $(40\pm2)\%$ at 20 $\g$.
 For $\pp$ mesons, an increase from $(21\pm2)\%$ to $(35\pm4)\%$ 
 is seen in the range from 5.5 to 20 $\g$. 

The proper decay length measurement allows us to
divide the data into two samples: a short-lived sample dominated by
prompt production, and a long-lived sample dominated by $B$-decays. 
 The short-lived sample is defined by
 $-0.1\leq ct \leq 0.013[0.01]$ cm, and the long-lived sample by
        $0.013[0.01]\leq ct \leq 0.3$ cm, for the $\jp$ $[\pp]$
analysis respectively. The boundary between the two $ct$ regions
has been optimized separately for the $\jp$ and $\pp$ samples, to 
maximize the purity of prompt decays in the short-lived sample and
$B$-decays in the long-lived sample.
 Depending on $P_T^{\psi}$,
 the prompt fraction in the short-lived sample ranges from 85\%[86\%]
 to 96\%[95\%],
and the $B$-decay fraction in the long-lived sample ranges from 83\%[86\%] to
98\%[91\%], for $\jp[\pp]$ respectively.
 
%%%%%%%%%%%%%% J/psi Results %%%%%%%%%%%%%%%%%%%%%%%%%%%%

The $\jp$ polarization is measured in seven $P_T$ 
bins, covering a range of $4-20 \ \g$. Using a 3 standard deviation
 mass window around the $\jp$ peak, our data sample consists of 
180,000 signal $\jp$ events, with a signal to background ratio of about 13.
The $\jp$ sample 
is divided into three subsamples: the short-lived and long-lived
SVX samples described above, and a third sample (the CTC sample) 
in which neither muon has SVX information and no $ct$ measurement is made.
 In each $\ptj$ bin, we measure the prompt
polarization ($\aP$) and the effective
 polarization of $\jp$ mesons from
$B$-hadron decays ($\ab$).
(We refer to $\ab$ as ``effective''
 because $\theta^{\ast}$ is defined using the
lab frame, not the $B$-hadron rest frame --- in effect this dilutes
 any polarization from the $B$-decay toward zero.)
We find that it is not feasible to make separate polarization 
measurements for direct $\jp$ production
and for production from $\chi_c$ and $\pp$ decays. The latter sources account
for 36$\pm$6\% of the prompt component, with only a small $\ptj$
 dependence \cite{cdfchi}.

The $\jp$ polarization is measured by fitting $\cts$ distributions
 in data to a set of Monte Carlo templates \cite{rjcthesis}.
The templates are generated by processing simulated samples of 
$\jp \rightarrow \mu^+ \mu^-$ decays  
with a detector and trigger simulation. The polarization is obtained using
a $\chi^2$ fit of the data to a weighted sum of transversely 
polarized and longitudinally polarized templates. The fitted weights 
 yield the polarization. 
Two transverse/longitudinal template pairs are generated, using measured
 prompt and $B$-decay $\ptj$ spectra \cite{cdfpsi}.
 The $\cts$ distribution of background events is modeled in the fit using  
sidebands around the $\jp$ mass peak.
The fit is performed simultaneously on
 the SVX short-lived, SVX long-lived and CTC samples, with two
fit parameters: $\aP$ and $\ab$. To account for the mixture of prompt
 and $B$-decay 
components in each sample, the relative fractions of prompt and $B$-decay
templates in each  
are fixed in the fit using the results of the lifetime fit.
 The B-decay fraction in the CTC sample is assumed to be the same as in
 the SVX sample, because the two samples
 differ primarily in the $z$ position of the primary vertex.
Within each $\ptj$ bin, a small correction is applied to the
 $\ptj$ distributions of the Monte Carlo samples so that they
match with those in the data. As an example, the fit in the $P_T$ range 
12-15 $\g$  is shown in Fig.\ \ref{Fjex}.

Three sources of systematic uncertainty are evaluated: the trigger
efficiency, the fitted prompt and $B$-decay fractions, and the $\ptj$
spectra used in making the Monte Carlo templates.
Except in the lowest
$P_T$ bins, the systematic uncertainties are much smaller than
  the statistical uncertainties.
Our fit results are listed in Table \ref{Tjp},
 and $\aP$ is compared with a theoretical
 NRQCD prediction \cite{braaten} in Fig.\ \ref{Fpol}.

%%%%%%%%%%%%%% Psi(2S) Results %%%%%%%%%%%%%%%%%%%%%%%%%%%%

The measurement of the $\pp$ polarization is made in three $P_T$
bins covering $5.5 - 20.0 \ \g$. Both muons are required to be
reconstructed in the SVX.  The resulting dimuon mass distribution is
fitted with a Gaussian signal and a linear background. We find a
total of $1855\pm65$ signal $\pp$ events, with a signal to
background ratio of about 1 in a $3$ standard deviation mass window
around the $\pp$ mass.

As discussed above, the sample in each $P_T$ bin is further divided
into two subsamples based on the $ct$ distribution. 
Because the statistics are lower than in the $\jp$ case,
we use 10 bins in $|\cts|$.
  The number of signal events in each $|\cts|$ bin is obtained by
fitting its mass distribution.  The resulting $|\cts|$ distributions
in the two $ct$ subsamples are fitted simultaneously to the
predicted number of events to extract the $\pp$ polarizations for
prompt and $B$-decay production.  The number of predicted events in
each $|\cts|$ bin is derived by weighting the normalized angular
distribution $I(\cts)$ with the detector acceptance \cite{pnthesis}.
 We use the measured
prompt and $B$-decay $P_T^{\psi(2S)}$ distributions\cite{cdfpsi} to
calculate the acceptance. As in the $\jp$ case, there is a small
correlation between the measured $P_T^{\psi(2S)}$ distributions
and the polarization. A correction is
applied iteratively in the fits to account for this dependence.  Figure
\ref{Fp2ex} shows the observed angular distributions with
their polarization fits for the short-lived sample in the three
$P_T^{\pp}$ bins.

Three sources of systematic uncertainty are considered: the
uncertainty in the event yield from the mass fits in the $|\cts|$
bins, the uncertainty due to the error on the fitted prompt and
$B$-decay fractions, and the uncertainty on the $|\cts|$ acceptance
from the Monte Carlo modeling of the $P_T^{\psi(2S)}$ distributions.
The uncertainty due to the trigger efficiency is negligible in the
$P_T^{\pp}$ range used. The systematic uncertainties are
much smaller than the statistical uncertainties.  The fitted values of
$\aP$ and $\ab$ as a function of $P_T^{\pp}$ are listed in Table
\ref{Tpp}, and $\aP$ is shown in Fig.\ \ref{Fpol} with
the NRQCD predictions \cite{benkram,braaten}.

%%%%%%%%%%%%%% Conclusion %%%%%%%%%%%%%%%%%%%%%%%%%%%%%%%

In conclusion, we have measured the polarization of 
 $\jp$ and $\pp$ mesons produced in 1.8 TeV $p\bar{p}$ collisions.
The polarization from $B$-decays is generally consistent with zero, as
expected. In both the $\jp$ and $\pp$
 cases, we do not observe increasing prompt transverse
 polarization at $P_T \gessim 12 \ \g$.
 Our measurements are
limited by statistics, especially for the $\pp$,
 but they appear to indicate that no large transverse
prompt polarization is present at high $P_T$, in disagreement with NRQCD
 factorization predictions.

%%%%%%%%%%%%%% Acknowledgements %%%%%%%%%%%%%%%%%%%%%%%%%%%%%%%%%%

We thank the Fermilab staff and the technical staffs of the
participating institutions for their vital contributions.  This work was
supported by the U.S. Department of Energy and National Science Foundation;
the Italian Istituto Nazionale di Fisica Nucleare; the Ministry of Education,
Science, Sports and Culture of Japan; the Natural Sciences and Engineering
Research
Council of Canada; the National Science Council of the Republic of China; 
the Swiss National Science Foundation; the A. P. Sloan Foundation; and the
Bundesministerium f\"{u}r Bildung und Forschung, Germany.

%%%%%%%%%%%%%% References %%%%%%%%%%%%%%%%%%%%%%%%%%%%%%%%%%%%%%%%

%%%%%%%%%%%%%%%%%%% Tables %%%%%%%%%%%%%%%%%%%%%%%%%%%%%%%%%%%%

\begin{table}
\caption{Fit results for $\jp$ polarization, with
statistical and systematic uncertainties.}
\label{Tjp}
\begin{tabular}{|c|c|c|c|} 
$P_T$ bin ($\g$) & Mean $P_T$  ($\g$) & $\aP$ & $\ab$ \\ \hline 
$4-5$   & 4.5  & $0.30\pm0.12\pm0.12$    & $-0.49\pm0.41\pm0.13$   \\ 
$5-6$   & 5.5  & $0.01\pm0.10\pm0.07$    & $-0.18\pm0.33\pm0.07$  \\ 
$6-8$   & 6.9  & $0.178\pm0.072\pm0.036$ & $0.10\pm0.20\pm0.04$  \\
$8-10$  & 8.8  & $0.323\pm0.094\pm0.019$ & $-0.06\pm0.20\pm0.02$  \\
$10-12$ & 10.8 & $0.26\pm0.14\pm0.02$    & $-0.19\pm0.23\pm0.02$  \\
$12-15$ & 13.2 & $0.11\pm0.17\pm0.01$    & $0.11\pm^{0.31}_{0.28}\pm0.02$  \\
$15-20$ & 16.7 & $-0.29\pm0.23\pm0.03$   & $-0.16\pm^{0.38}_{0.33}\pm0.05$  \\ 
\end{tabular}
\end{table}

\begin{table}
\caption{Fit results for $\pp$ polarization, with
statistical and systematic uncertainties.}
\label{Tpp}
\begin{tabular}{|c|c|c|c|} 
$P_T$ bin ($\g$) & Mean $P_T$  ($\g$) & $\aP$ & $\ab$ \\ \hline
$5.5-7.0$  & 6.2  & $-0.08\pm0.63\pm0.02$ & $-0.26\pm1.26\pm0.04$ \\
$7.0-9.0$  & 7.9  & $ 0.50\pm0.76\pm0.04$ & $-1.68\pm0.55\pm0.12$ \\
$9.0-20.0$ & 11.6 & $-0.54\pm0.48\pm0.04$ & $ 0.27\pm0.81\pm0.06$ \\ 
\end{tabular}
\end{table}

%%%%%%%%%%%%%%%%%%%% Figures %%%%%%%%%%%%%%%%%%%%%%%%%%%%%%%%%

\begin{center} % for Draft

\begin{figure}
\epsfig{file=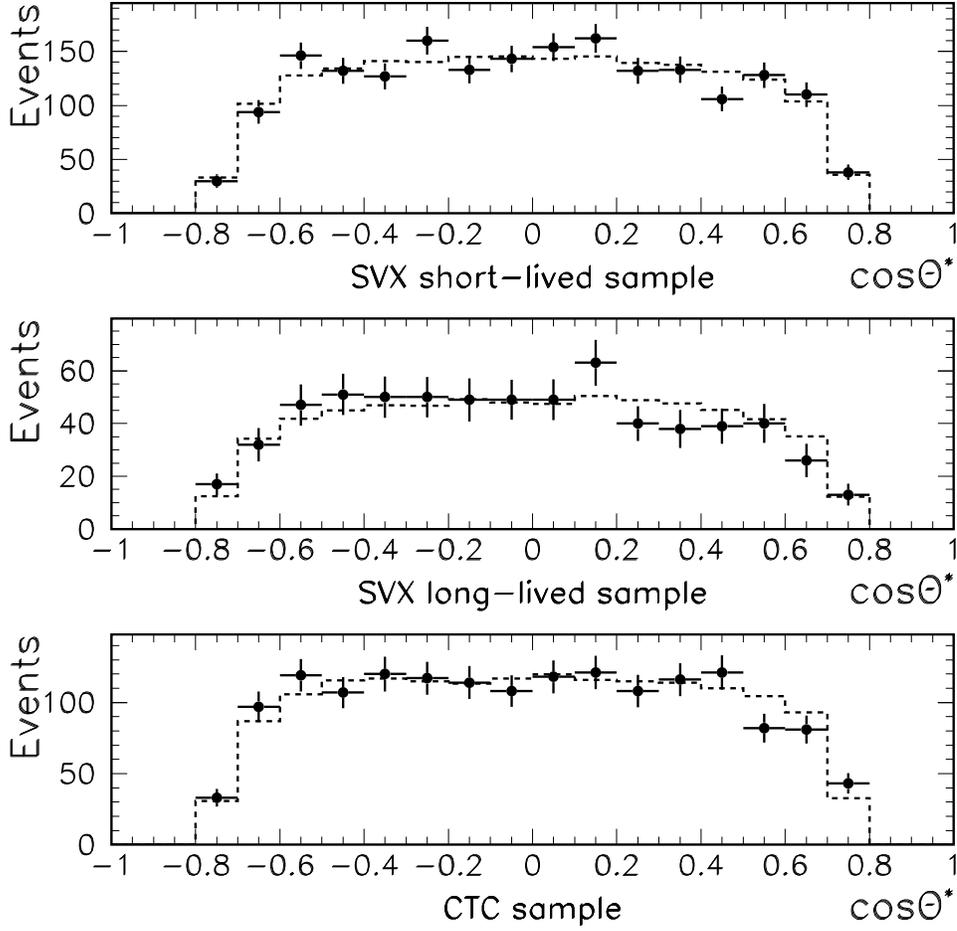,width=14cm} % for Draft
\caption{The $\jp$ polarization fit to $\cts$ distributions in
 the 12-15 $\g$ bin. Points: sideband-subtracted data in the
 SVX short-lived, SVX long-lived and CTC samples. 
 Dashed lines: fit.}
\label{Fjex}
\end{figure}

\begin{figure}
\epsfig{file=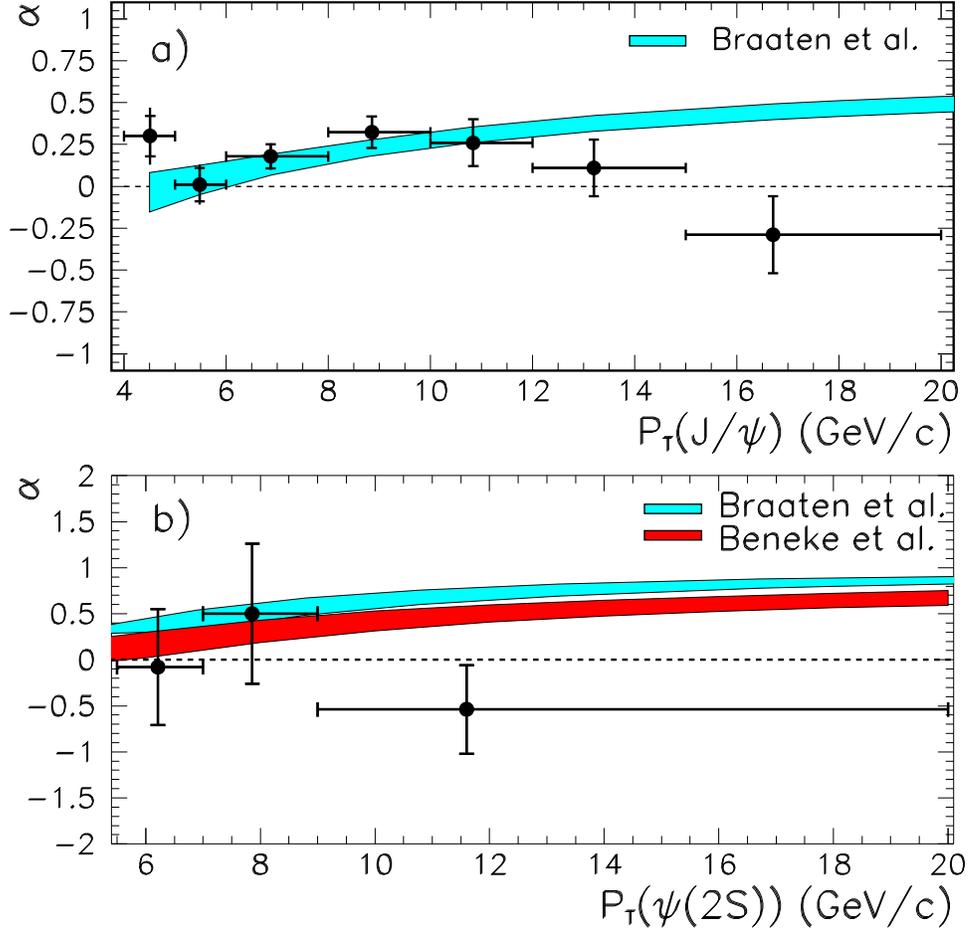,width=14cm} % for Draft
\caption{(a) The fitted polarization of prompt $\jp$ mesons 
 for $|y^{\jp}| < 0.6$.
 Full error bars denote 
statistical and systematic uncertainties added in quadrature; ticks
denote statistical errors alone.
 The shaded band shows an
 NRQCD factorization prediction \protect\cite{braaten} which 
includes the contribution from $\chi_c$ and $\pp$ decays. 
(b) The fitted polarization of prompt $\pp$ mesons for $|y^{\pp}| < 0.6$.
 Error bars denote statistical
and systematic uncertainties added in quadrature.
Shaded bands show two NRQCD factorization predictions
\protect\cite{benkram,braaten}. }
\label{Fpol}
\end{figure}

\begin{figure}
\epsfig{file=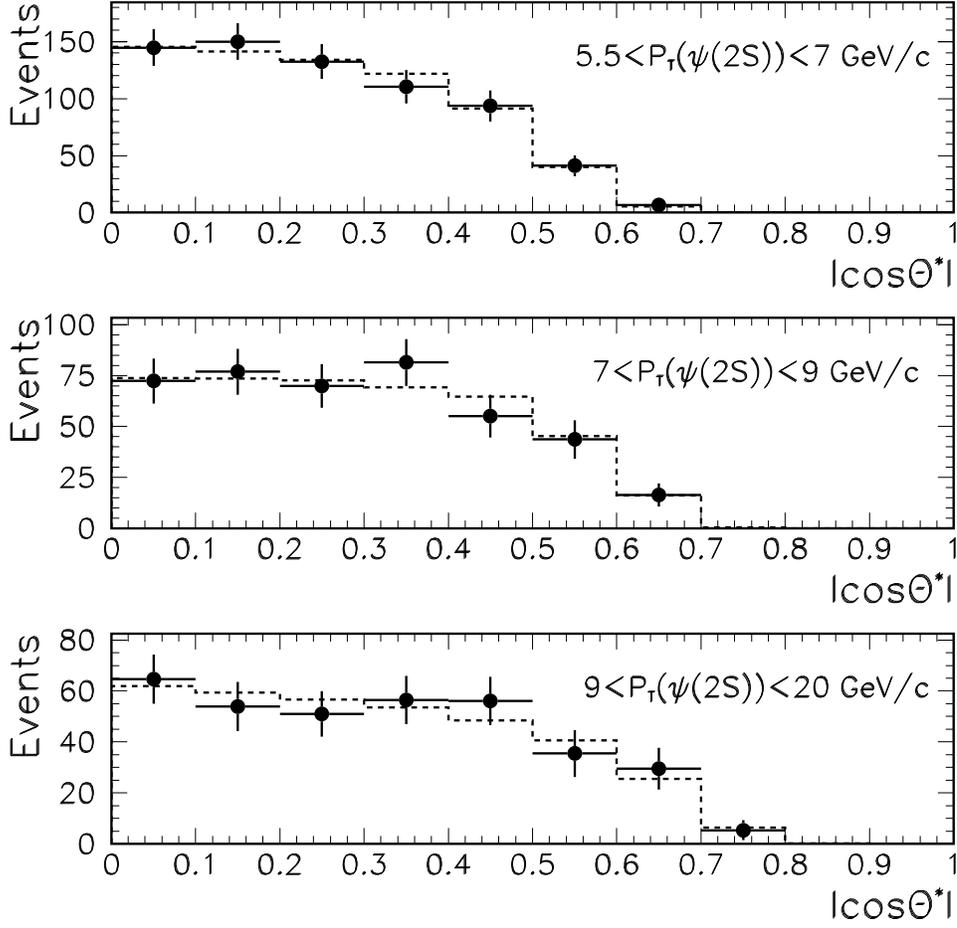,width=14cm} % for Draft
\caption{Fits to $|\cts|$ distributions in the short-lived $\pp$ data
sample, in the three $P_T$ bins. Points: data, 
dashed lines: fit. The acceptance extends further out in $|\cts|$ as
$P_T$ increases.}
\label{Fp2ex}
\end{figure}

\end{center} % for Draft

\end{document}

%% file: cdfalist.tex
\font\eightit=cmti8
\def\r#1{\ignorespaces $^{#1}$}
\hfilneg
\begin{sloppypar}
\noindent
T.~Affolder,\r {21} H.~Akimoto,\r {43}
A.~Akopian,\r {36} M.~G.~Albrow,\r {10} P.~Amaral,\r 7 S.~R.~Amendolia,\r {32} 
D.~Amidei,\r {24} K.~Anikeev,\r {22} J.~Antos,\r 1 
G.~Apollinari,\r {10}
% ...
\end{sloppypar}
\vskip .026in
\begin{center}
(CDF Collaboration)
\end{center}

\vskip .026in
\begin{center}
\r 1  {\eightit Institute of Physics, Academia Sinica, Taipei, Taiwan 11529, 
Republic of China} \\
\r 2  {\eightit Argonne National Laboratory, Argonne, Illinois 60439} \\
\r 3  {\eightit Istituto Nazionale di Fisica Nucleare, University of Bologna,
I-40127 Bologna, Italy} \\
\r 4  {\eightit Brandeis University, Waltham, Massachusetts 02254} \\
\r 5  {\eightit University of California at Los Angeles, Los 
Angeles, California  90024} \\  
\r 6  {\eightit Instituto de Fisica de Cantabria, CSIC-University of Cantabria, 
39005 Santander, Spain} \\
\r 7  {\eightit Enrico Fermi Institute, University of Chicago, Chicago, 
Illinois 60637} \\
\r 8  {\eightit Joint Institute for Nuclear Research, RU-141980 Dubna, Russia}
\\
\r 9  {\eightit Duke University, Durham, North Carolina  27708} \\
\r {10}  {\eightit Fermi National Accelerator Laboratory, Batavia, Illinois 
60510} \\
\r {11} {\eightit University of Florida, Gainesville, Florida  32611} \\
\r {12} {\eightit Laboratori Nazionali di Frascati, Istituto Nazionale di Fisica
               Nucleare, I-00044 Frascati, Italy} \\
\r {13} {\eightit University of Geneva, CH-1211 Geneva 4, Switzerland} \\
\r {14} {\eightit Harvard University, Cambridge, Massachusetts 02138} \\
\r {15} {\eightit Hiroshima University, Higashi-Hiroshima 724, Japan} \\
\r {16} {\eightit University of Illinois, Urbana, Illinois 61801} \\
\r {17} {\eightit The Johns Hopkins University, Baltimore, Maryland 21218} \\
\r {18} {\eightit Institut f\"{u}r Experimentelle Kernphysik, 
Universit\"{a}t Karlsruhe, 76128 Karlsruhe, Germany} \\
\r {19} {\eightit Korean Hadron Collider Laboratory: Kyungpook National
University, Taegu 702-701; Seoul National University, Seoul 151-742; and
SungKyunKwan University, Suwon 440-746; Korea} \\
\r {20} {\eightit High Energy Accelerator Research Organization (KEK), Tsukuba, 
Ibaraki 305, Japan} \\
\r {21} {\eightit Ernest Orlando Lawrence Berkeley National Laboratory, 
Berkeley, California 94720} \\
\r {22} {\eightit Massachusetts Institute of Technology, Cambridge,
Massachusetts  02139} \\   
\r {23} {\eightit Institute of Particle Physics: McGill University, Montreal 
H3A 2T8; and University of Toronto, Toronto M5S 1A7; Canada} \\
\r {24} {\eightit University of Michigan, Ann Arbor, Michigan 48109} \\
\r {25} {\eightit Michigan State University, East Lansing, Michigan  48824} \\
\r {26} {\eightit University of New Mexico, Albuquerque, New Mexico 87131} \\
\r {27} {\eightit The Ohio State University, Columbus, Ohio  43210} \\
\r {28} {\eightit Osaka City University, Osaka 588, Japan} \\
\r {29} {\eightit University of Oxford, Oxford OX1 3RH, United Kingdom} \\
\r {30} {\eightit Universita di Padova, Istituto Nazionale di Fisica 
          Nucleare, Sezione di Padova, I-35131 Padova, Italy} \\
\r {31} {\eightit University of Pennsylvania, Philadelphia, 
        Pennsylvania 19104} \\   
\r {32} {\eightit Istituto Nazionale di Fisica Nucleare, University and Scuola
               Normale Superiore of Pisa, I-56100 Pisa, Italy} \\
\r {33} {\eightit University of Pittsburgh, Pittsburgh, Pennsylvania 15260} \\
\r {34} {\eightit Purdue University, West Lafayette, Indiana 47907} \\
\r {35} {\eightit University of Rochester, Rochester, New York 14627} \\
\r {36} {\eightit Rockefeller University, New York, New York 10021} \\
\r {37} {\eightit Rutgers University, Piscataway, New Jersey 08855} \\
\r {38} {\eightit Texas A\&M University, College Station, Texas 77843} \\
\r {39} {\eightit Texas Tech University, Lubbock, Texas 79409} \\
\r {40} {\eightit Istituto Nazionale di Fisica Nucleare, University of Trieste/
Udine, Italy} \\
\r {41} {\eightit University of Tsukuba, Tsukuba, Ibaraki 305, Japan} \\
\r {42} {\eightit Tufts University, Medford, Massachusetts 02155} \\
\r {43} {\eightit Waseda University, Tokyo 169, Japan} \\
\r {44} {\eightit University of Wisconsin, Madison, Wisconsin 53706} \\
\r {45} {\eightit Yale University, New Haven, Connecticut 06520} \\
\end{center}